\documentclass[submission, Proceedings]{SciPost}

\hypersetup{
    colorlinks,
    linkcolor={red!50!black},
    citecolor={blue!50!black},
    urlcolor={blue!80!black}
}

\usepackage[bitstream-charter]{mathdesign}
\DeclareSymbolFont{usualmathcal}{OMS}{cmsy}{m}{n}
\DeclareSymbolFontAlphabet{\mathcal}{usualmathcal}

\usepackage{amsmath}
\usepackage{amsfonts}
\usepackage{graphicx}
\usepackage{caption}
\usepackage{subcaption}
\usepackage{float}

\newcommand{\pT}{\ensuremath{p_T}}
\newcommand{\pizero}{\ensuremath{\pi^0}}
\newcommand{\ALL}{\ensuremath{A_{LL}}}

\begin{document}
\graphicspath{{../}{../../spd2021/}{../../prd-102-032001/}}

\begin{center}{\Large \textbf{
Longitudinal double helicity asymmetry $A_{LL}$ from direct photon, jet and charged pion production in polarized $\vec{p}+\vec{p}$ collisions\\
}}\end{center}

\begin{center}
Zhongling Ji\textsuperscript{1$\star$} for the PHENIX Collaboration
\end{center}

\begin{center}
{\bf 1} Stony Brook University
\\
* zhongling.ji@stonybrook.edu
\end{center}

\begin{center}
\today
\end{center}


\definecolor{palegray}{gray}{0.95}
\begin{center}
\colorbox{palegray}{
  \begin{tabular}{rr}
  \begin{minipage}{0.1\textwidth}
    \includegraphics[width=22mm]{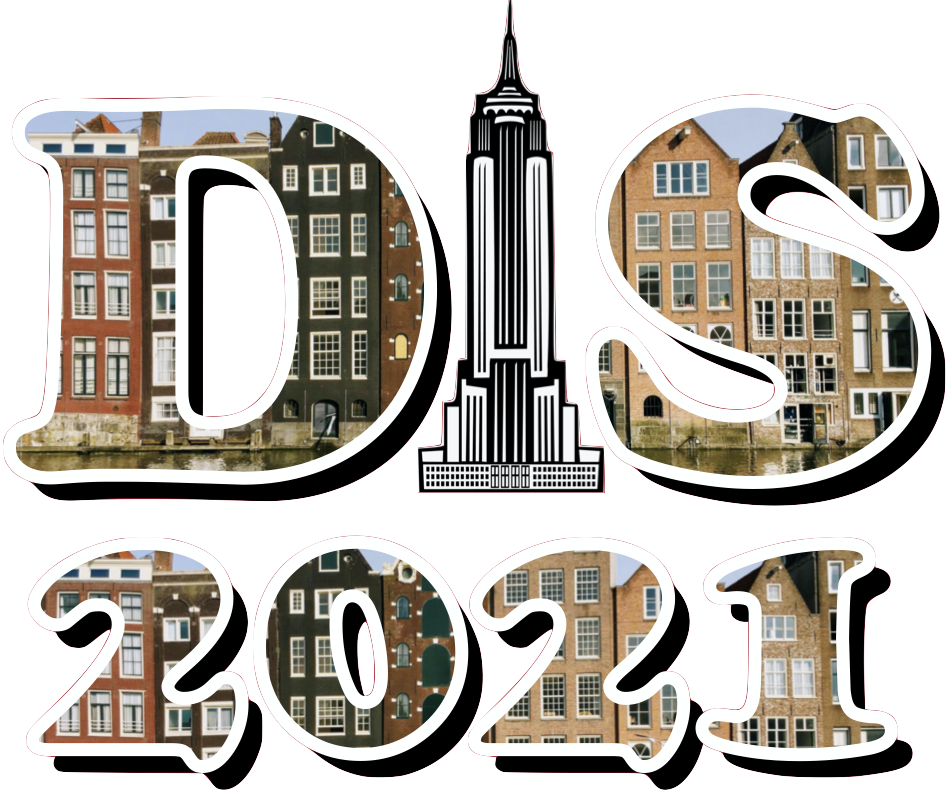}
  \end{minipage}
  &
  \begin{minipage}{0.75\textwidth}
    \begin{center}
    {\it Proceedings for the XXVIII International Workshop\\ on Deep-Inelastic Scattering and
Related Subjects,}\\
    {\it Stony Brook University, New York, USA, 12-16 April 2021} \\
    \doi{10.21468/SciPostPhysProc.?}\\
    \end{center}
  \end{minipage}
\end{tabular}
}
\end{center}

\section*{Abstract}
{\bf
Understanding the proton spin composition from the quarks and gluons spin polarization and their motion is important to test various kinds of sum rules and nonperturbative properties of hadrons. At the Relativistic Heavy Ion Collider (RHIC), we collide longitudinally polarized proton beams and measure the double helicity asymmetry $A_{LL}$, which is an important physical quantity for extracting the polarized parton distribution functions (PDFs) of the proton. Direct photon, jet and charged pion production are good channels to probe the gluon spin polarization inside the proton, with the ability to probe also the sign of the gluon spin. Direct photon production is the theoretically ``cleanest'' channel, with little fragmentation contribution, but limited by statistics. On the other hand, jet and charged pion production have more statistics, but include more hard processes and hadronization effects. I will present the recent measurements of direct photon, jet and charged pion $A_{LL}$s at PHENIX and show their complementary roles in extracting the gluon spin.
}

\vspace{10pt}
\noindent\rule{\textwidth}{1pt}
\tableofcontents\thispagestyle{fancy}
\noindent\rule{\textwidth}{1pt}
\vspace{10pt}

\section{Introduction}

Nucleons (protons and neutrons) and electrons are the building block of the ordinary matters. While electrons are taken as fundamental particles with point structure in the Standard Model (SM), protons and neutrons are composite particles with rich structures. These structures are usually revealed through high energy collisions and take partons (quarks and gluons) as the underlying degrees of freedom. One of the interesting structures is the longitudinal spin decomposition of the proton, which decomposes the proton spin as the spin and orbital momentum contributions from quarks and gluons
\begin{equation}
\frac{1}{2} = \frac{1}{2} \Delta\Sigma + \Delta G + L_q + L_g,
\end{equation}
where $\frac{1}{2}\Delta\Sigma$ and $\Delta G$ are spin contributions from quarks and gluons, respectively, $L_q$ and $L_g$ are their orbital momentum contributions. The spin contributions can further be expressed based on longitudinal momentum fractions of partons
\begin{align}
\Delta\Sigma =& \sum\limits_{q=u,d,s} \int_0^1 \Delta q(x)dx, \\
\Delta G =& \int_0^1 \Delta g(x)dx,
\end{align}
where $x = p^+/P^+$ is the fraction of longitudinal momentum carried by the parton and $P^+$ ($p^+$) is the proton's (parton's) momentum in the light cone frame.

The polarized inclusive deep inelastic scattering (pDIS) \cite{1988364, ASHMAN19891, ALEXAKHIN20078, PhysRevD.75.012007} probes the singlet axial charge $g_A^{(0)}$, which is related to parton spin
\begin{equation}
g_A^{(0)}|_{\text{pDIS}} = \Delta\Sigma - 3\frac{\alpha_s}{2\pi}\Delta G,
\end{equation}
where $\alpha_s$ is the strong coupling constant. The polarized semi-inclusive deep inelastic scattering (pSIDIS) probes the gluon spin $\Delta g(x)$ through next-to-leading order (NLO) photon-gluon fusion process $\gamma^*g\rightarrow q\bar{q}$ \cite{PhysRevD.70.012002, Airapetian2010}. The longitudinal polarized proton-proton collisions can probe the gluon spin at leading order (LO). Currently, the Relativistic Heavy Ion Collider (RHIC) is the only machine capable of colliding polarized proton beams \cite{ZELENSKI2005248}. Measurements of double helicity asymmetry (\ALL) of direct photon, jet and charged pion productions in polarized $\vec{p}+\vec{p}$ collisions are three complementary channels to probe the gluon spin inside the proton. Direct photon with an isolation criterion is mainly produced by quark-gluon Compton scattering $qg\rightarrow q\gamma$. Since there is little fragmentation contribution to the direct photon production, this is the ``cleanest'' channel. But it is limited by statistics due to the suppression by the electromagnetic coupling constant. On the other hand, the jet and charged pion productions probe the gluon spin through $qg\rightarrow qg$ and $gg\rightarrow gg$ scatterings, which have more statistics, but are influenced more by hadronization processes. In addition, the charged pion productions have the ability to separate the $u$ and $d$ quark contributions. Recent RHIC measurements of \pizero\ and jets at  $\sqrt{s}$ = 62.4 and 200 GeV \cite{PhysRevD.90.012007, PhysRevLett.103.012003, PhysRevD.79.012003, PhysRevD.86.032006, PhysRevLett.115.092002} included in global analyses have shown the first direct evidence of nonzero gluon spin contributions to the spin of the proton \cite{PhysRevLett.113.012001, 2014276} in the Bjorken-$x$ range larger than 0.05. Measurements at higher energy $\sqrt{s}$ = 510 GeV \cite{PhysRevD.93.011501, PhysRevD.100.052005} have confirmed the nonzero gluon polarization and extended the minimum $x$ reach to $\sim$0.01. I will present the recent PHENIX spin measurements from direct photon, jet and charged pion.

\section{Experimental setup}

The measurements of direct photon, jet and charged pion \ALL\ are taken at $\sqrt{s}$ = 510 GeV from the 2013 RHIC running period by the PHENIX detector \cite{ADCOX2003469} in midrapidity, which has pseudorapidy coverage $|\eta| <$ 0.35 and $\pi$ coverage for azimuthal angle $\phi$. The primary detector for high energy photons is an electromagnetic calorimeter (EMCal) \cite{APHECETCHE2003521}, consisting of two subsystems, a six sector lead-scintillator (PbSc), and a two sector lead glass (PbGl) detector, each located 5 m radially from the beam line. The EMCal has fine granularity with each tower covering $\Delta\eta \times \Delta\phi \sim$ 0.01 $\times$ 0.01 (0.008 $\times$ 0.008) for PbSc (PbGl). The high energy photons are selected by an EMCal trigger, which combines each 4 $\times$ 4 towers of EMCal to a single module and requires an energy threshold ranging from 3.7 to 5.6 GeV. The Drift Chamber (DC) \cite{ADCOX2003489} is used to measure the momentum of charged particles in jet and charged pions and veto charged particles in direct photon measurements. The Pad Chamber 3 (PC3) is used to match the DC track in the jet measurement. The Ring \v{C}erenkov detector (RICH) \cite{AKIBA1999143} is used for particle identification (PID) of charged pions. The Beam-Beam Counters (BBC) \cite{ALLEN2003549} cover 3.1 $< |\eta| <$ 3.9 and are located at $\pm$144 cm from the interaction point along the beam line. The BBCs measure the collision vertex and provide a minimum-bias (MB) trigger. The BBCs are also used as a luminosity monitor.

The observed physical quantity is the double helicity asymmetry \ALL, which is defined as
\begin{equation}
A_{LL} = \frac{\Delta\sigma}{\sigma} = \frac{\sigma_{++}-\sigma_{+-}}{\sigma_{++}+\sigma_{+-}},
\end{equation}
where $\sigma_{++}$ ($\sigma_{+-}$) is the cross section for the same (opposite) helicity proton collisions. This can be rewritten in terms of particle yield and beam polarizations:
\begin{equation}
A_{LL} = \frac{1}{P_BP_Y} \frac{N_{++}-RN_{+-}}{N_{++}+RN_{+-}}
\end{equation}
where $N_{++}$ ($N_{+-}$) is the number of measured events from the collisions with the same (opposite) helicities. $P_{B}$ and $P_{Y}$ are the polarizations for the two proton beams, and the average values in 2013 were 0.55 and 0.57, respectively \cite{POBLAGUEV2020164261}. $R$ ($= \mathcal{L_{++}}/\mathcal{L_{+-}}$) is the relative luminosity that is measured by the BBC.

\section{Direct photon \ALL}

The sources of direct photon signal at leading order include: quark gluon Compton scattering $gq\rightarrow q\gamma$, quark antiquark annihilation $q\bar{q}\rightarrow g\gamma$, parton fragmentation to photon, and quark bremsstrahlung. The background photons mainly come from decay photons from hadrons, such as \pizero, $\eta$, $\omega$ and $\eta'$. The photons are required to pass the EMcal trigger and measured by the EMCal. The \pizero\ can be reconstructed by the invariant mass of its two decay photons in the mass range 110 $< M_{\gamma\gamma} <$ 160 MeV/c$^2$ ($M_{\pi^0} \pm 3\sigma$). The production ratios of $\eta$, $\omega$, $\eta'$ over \pizero\ are obtained based on the previous $\sqrt{s}$ = 200 GeV measurement \cite{PhysRevD.83.052004}. To reduce the hadron decay background and the photons from parton fragmentation and quark bremsstrahlung, an isolation criteria is used. For any other particles within a cone of radius $r_{cone} = \sqrt{(\delta\eta)^2 + (\delta\phi)^2} = 0.5$ of the signal photon, the isolation criteria requires the energy sum of neutral particles in the EMCal plus the momentum sum of charged particles in the DC is less than 10\% of the energy of the signal photon: $E_{cone} = E_{neutral} + p_{charged}\cdot c < 0.1 E_{\gamma}$.

\begin{figure} 
\centering
\includegraphics[width=0.5\textwidth]{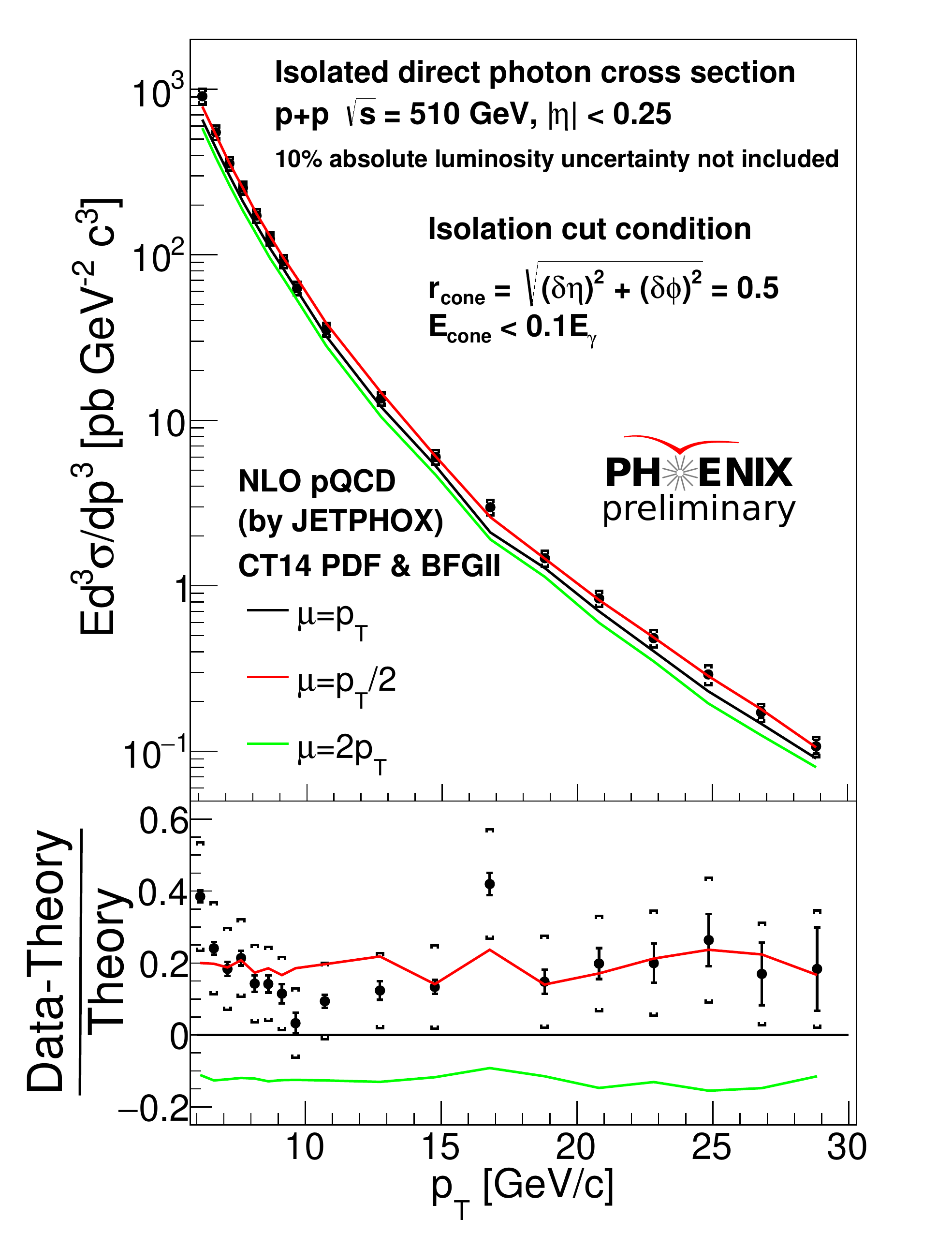}
\caption{Isolated direct photon cross section as a function of \pT\ compared with JETPHOX NLO pQCD calculations \cite{Catani_2002} for different renormalization and factorization scales $\mu$ = \pT/2 (dashed line), \pT\ (solid line), 2\pT\ (dotted line). The bars represent statistical uncertainties and square brackets are for systematic uncertainties. The bottom plot shows the comparison of data and calculations.}
\label{fig:iso}
\end{figure}

The isolated direct photon cross section at $|\eta| <$ 0.25 in psudorapidity at $\sqrt{s}$ = 510 GeV is shown in Fig.~\ref{fig:iso} as a function of \pT\ and compared with the JETPHOX NLO perturbative Quantum Chromodynamics (pQCD) calculation \cite{Catani_2002} using CT14 parton distribution functions (PDF) \cite{PhysRevD.93.033006} and BFGII fragmentation function (FF) \cite{Klasen2014}. The calculation is in good agreement with our data within the uncertainties.

\begin{figure}
\centering
\includegraphics[width=0.5\textwidth]{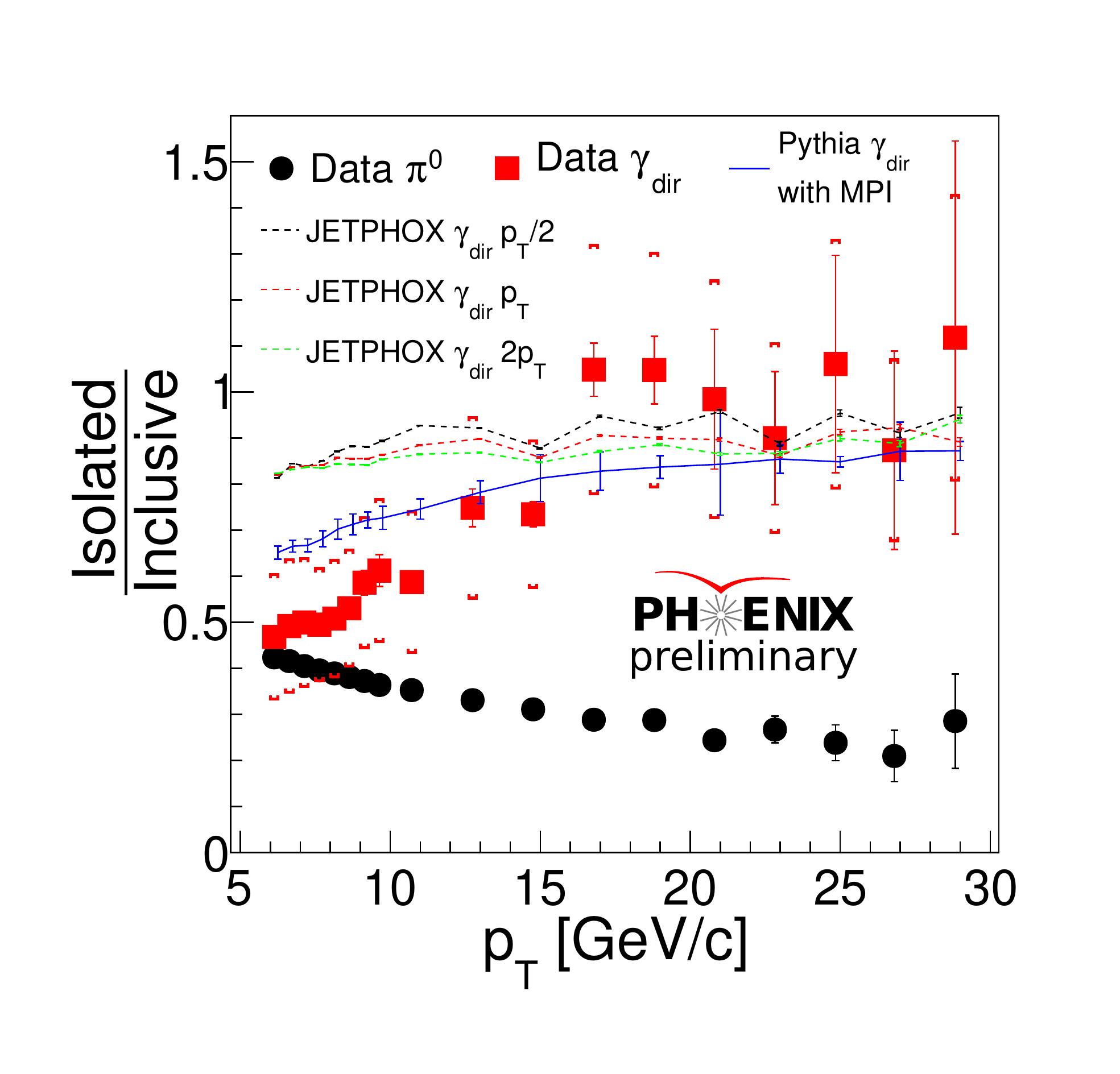}
\caption{Isolated over inclusive direct photon ratio: bars are statistical uncertainties and square brackets are systematic uncertainties. POWHEG + PYTHIA8 without MPI and JETPHOX NLO pQCD calculations with different renormalization and factorization scales are also shown.}
\label{fig:iso2inc}
\end{figure}

In Fig.~\ref{fig:iso2inc}, the ratio of isolated over inclusive direct photons is shown together with theoretical calculations. The MC simulations are from POWHEG + PYTHIA8 \cite{Nason_2004, Frixione_2007, Alioli2010, Jezo2016, Klasen2018}. POWHEG is a NLO partonic level generator, the output of which can be used as the input for PYTHIA8. The PYTHIA8 includes the multiparton interactions (MPI), parton showers (PS) and fragmentation processes. POWHEG uses CT14 PDF \cite{PhysRevD.93.033006}. There is no final-state factorization scale in PYTHIA8 as it uses string fragmentation instead. In the POWHEG+PYTHIA8 simulations, the uncertainties from renormalization and factorization scales are canceled between the numerator and denominator of the fraction. The JETPHOX NLO pQCD calculations overestimate the data due to its partonic nature. POWHEG+PYTHIA8 with MPI+PS gives better prediction for the isolated over inclusive direct photon cross section ratio.

\begin{figure}
\centering
\includegraphics[width=0.5\textwidth]{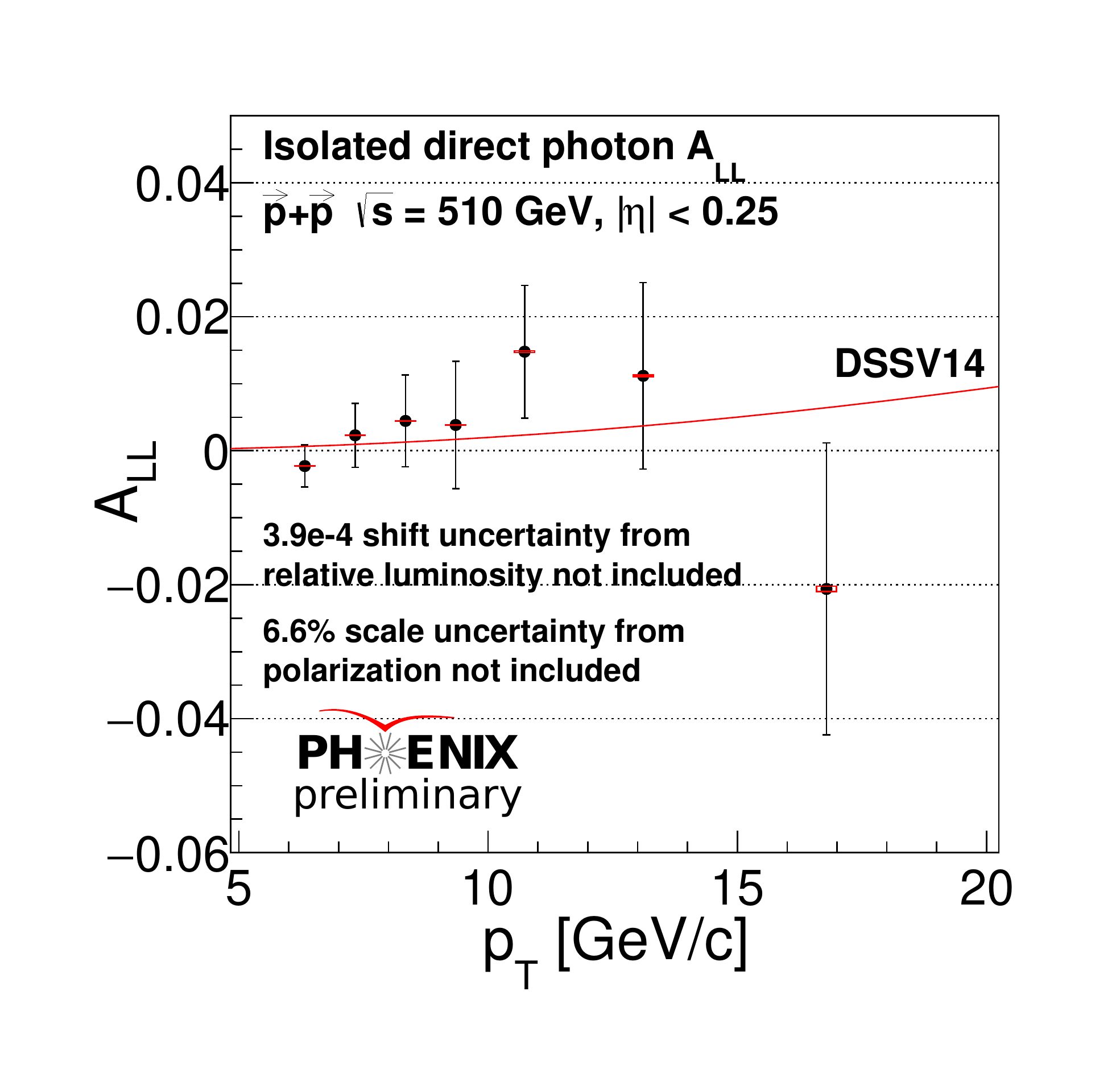}
\caption{Double helicity asymmetry $A_{LL}$ $vs$ $p_{T}$ for isolated direct photon production in polarized p+p collisions at $\sqrt{s}$ = 510 GeV at midrapidity. Vertical error bars (boxes) represent the statistical (systematic) uncertainties. The NLO pQCD calculation at $\sqrt{s}$ = 500 GeV is plotted as the solid curve.}
\label{fig:all}
\end{figure}

The double helicity asymmetry of isolated direct photon production in longitudinally polarized proton collisions at $|\eta| <$ 0.25 in psudorapidity at $\sqrt{s}$ = 510 GeV is shown in Fig.~\ref{fig:all} for 6 $< p_{T} <$ 20 GeV/c. The NLO pQCD calculation was obtained using DSSV14 polarized PDF, CT10 unpolarized PDF and GRV FF for the renormalization and factorization scales $\mu = p_T$ \cite{PhysRevLett.101.072001,PhysRevLett.113.012001,PhysRevD.100.114027}. The calculation is in good agreement with our results within the experimental uncertainties. This result will provide new constraint on gluon spin $\Delta G$ when included in the global analysis in the future.

\section{Jet \ALL}

The jet measurements use the anti-$k_T$ algorithm \cite{Cacciari_2008} with cone size R = 0.3 at $|\eta| <$ 0.15 in psudorapidity at $\sqrt{s}$ = 510 GeV. The momentum of charged particles is measured by the DC with a confirming match in PC3, while the energy of neutral particles is measured by the EMCal with a time-of-flight requirement to remove clusters from previous crossings due to the electronic device latency of EMCal. Good tracks and clusters are associated to avoid double counting. The selected events are required to pass the EMCal trigger with an energy threshold of 5.6 GeV. Only the jet with the largest \pT\ in the event is kept. The reconstructed \pT\ is unfolded to the true \pT\ by Bayes iteration method \cite{DAgostini:2010hil} implemented by RooUnfold.

\begin{figure} 
\centering
\includegraphics[width=0.5\textwidth]{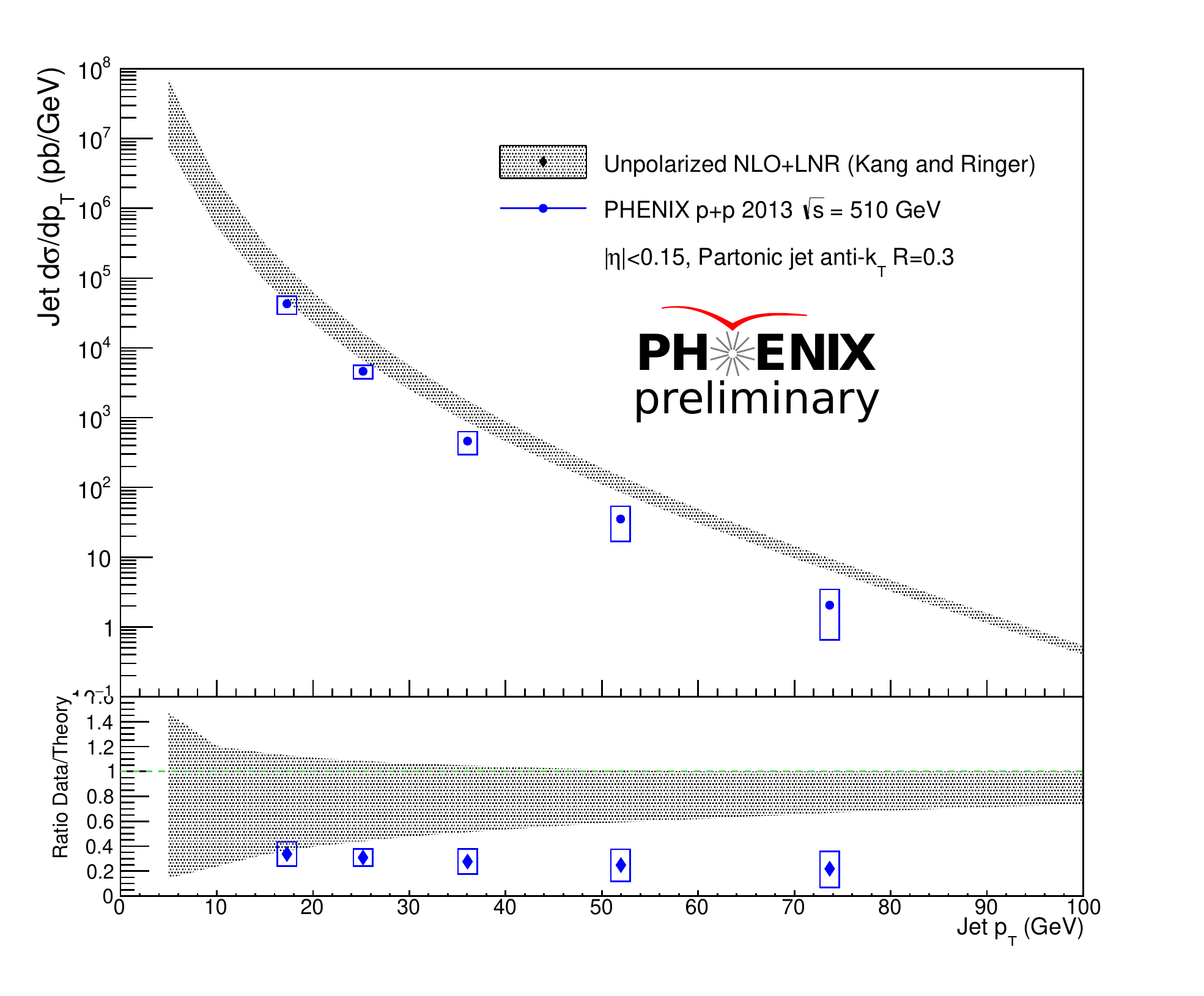}
\caption{Jet cross section as a function of \pT\ compared with NLO pQCD calculations with ln(R) resummations. The bars represent statistical uncertainties and square brackets are for systematic uncertainties. The bottom plot shows the comparison of data and calculations.}
\label{fig:xsec-jet}
\end{figure}

Fig.~\ref{fig:xsec-jet} shows the measured jet cross section and comparison with NLO pQCD calculation with ln(R) resummations. The calculation largely overestimates the data. Such a behavior is seen also by CMS when having small R sizes \cite{PhysRevLett.107.132001}.

\begin{figure} 
\centering
\includegraphics[width=0.5\textwidth]{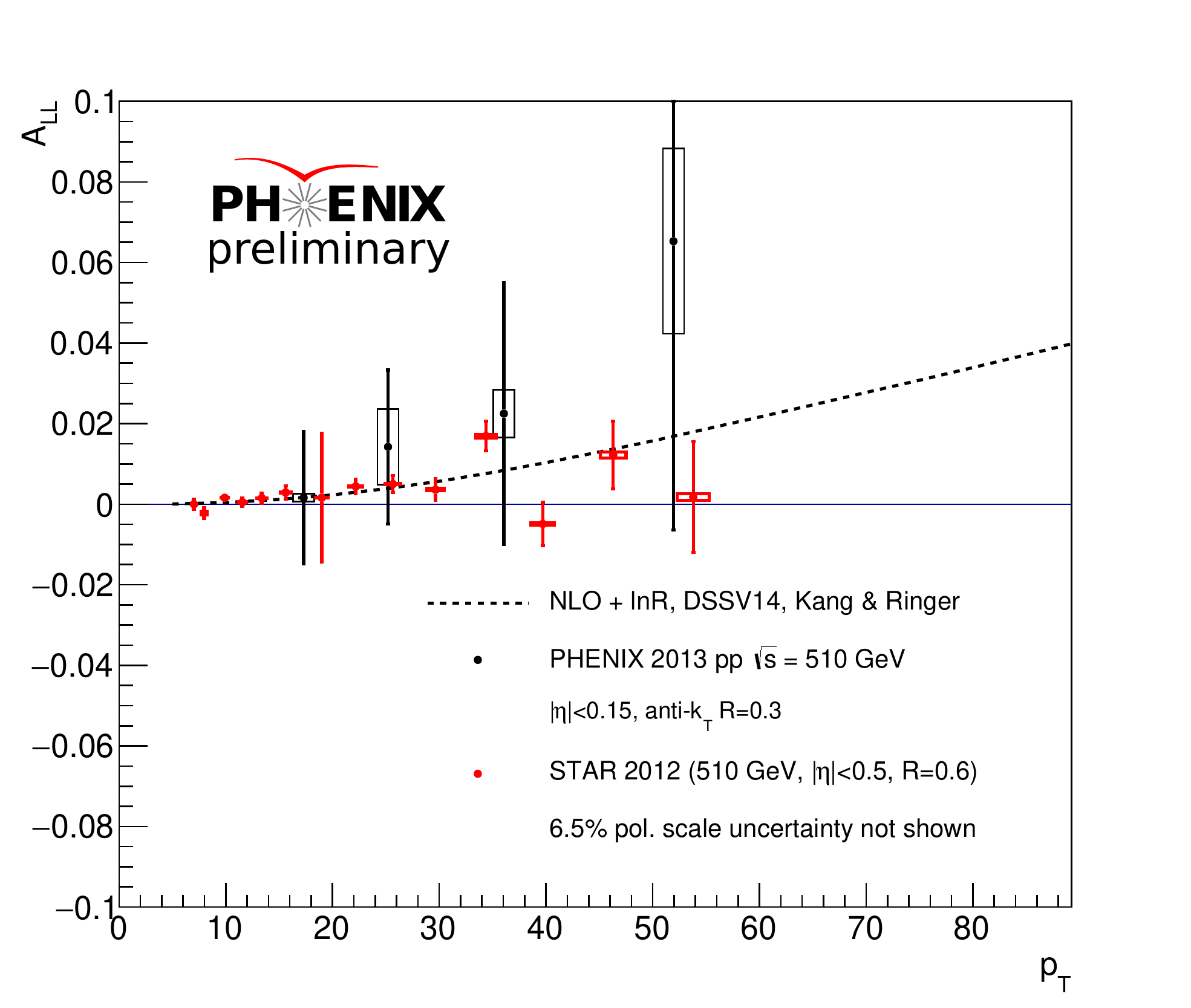}
\caption{Jet \ALL\ as a function of \pT\ compared with NLO pQCD calculations with ln(R) resummations. The bars represent statistical uncertainties and square brackets are for systematic uncertainties. The measurement from RHIC run 2012 at STAR is shown as a comparison.}
\label{fig:all-jet}
\end{figure}

Fig.~\ref{fig:all-jet} shows the jet \ALL\ measurement at PHENIX by using data from RHIC run 2013. The systematic uncertainties are correlated due to unfolding. The STAR jet \ALL\ measurement from RHIC run 2012 is also shown as a comparison. The theory calculation from NLO pQCD with ln(R) resummations and DSSV14 polarized PDF \cite{PhysRevLett.113.012001} is consistent with data. These measurements will provide independent constraint on polarized gluon PDF $\Delta g(x)$ when included in the global analysis in the future.

\section{Charged pion \ALL}

The charged pions are measured at PHENIX at $|\eta| <$ 0.35 in psudorapidity at $\sqrt{s}$ = 510 GeV \cite{PhysRevD.102.032001}. The particles are required to pass a preselection rule with cluster energy over reconstructed momentum ratio 0.2 $< E/p <$ 0.8, since the signal-to-background ratio is high in this range (fig.~4 of \cite{PhysRevD.102.032001}). The preselected particles are required to fire the EMCal trigger. Only the small fraction of pions that shower already in the EMCAL get selected with this trigger. In addition to the trigger, the charged pion candidate track measured by the DC is required to be pointing to the EMCal tower that fired the trigger. RICH is used as an additional PID and more than one photomultiplier is required to be fired by the charged pion candidate with threshold of $\sim$4.9 GeV. The next threshold of RICH is 17.3 GeV from kaons. An EM shower shape requirement is also used to reduce electron background.

\begin{figure}
\centering
\begin{subfigure}[t]{0.48\textwidth}
\includegraphics[width=\textwidth]{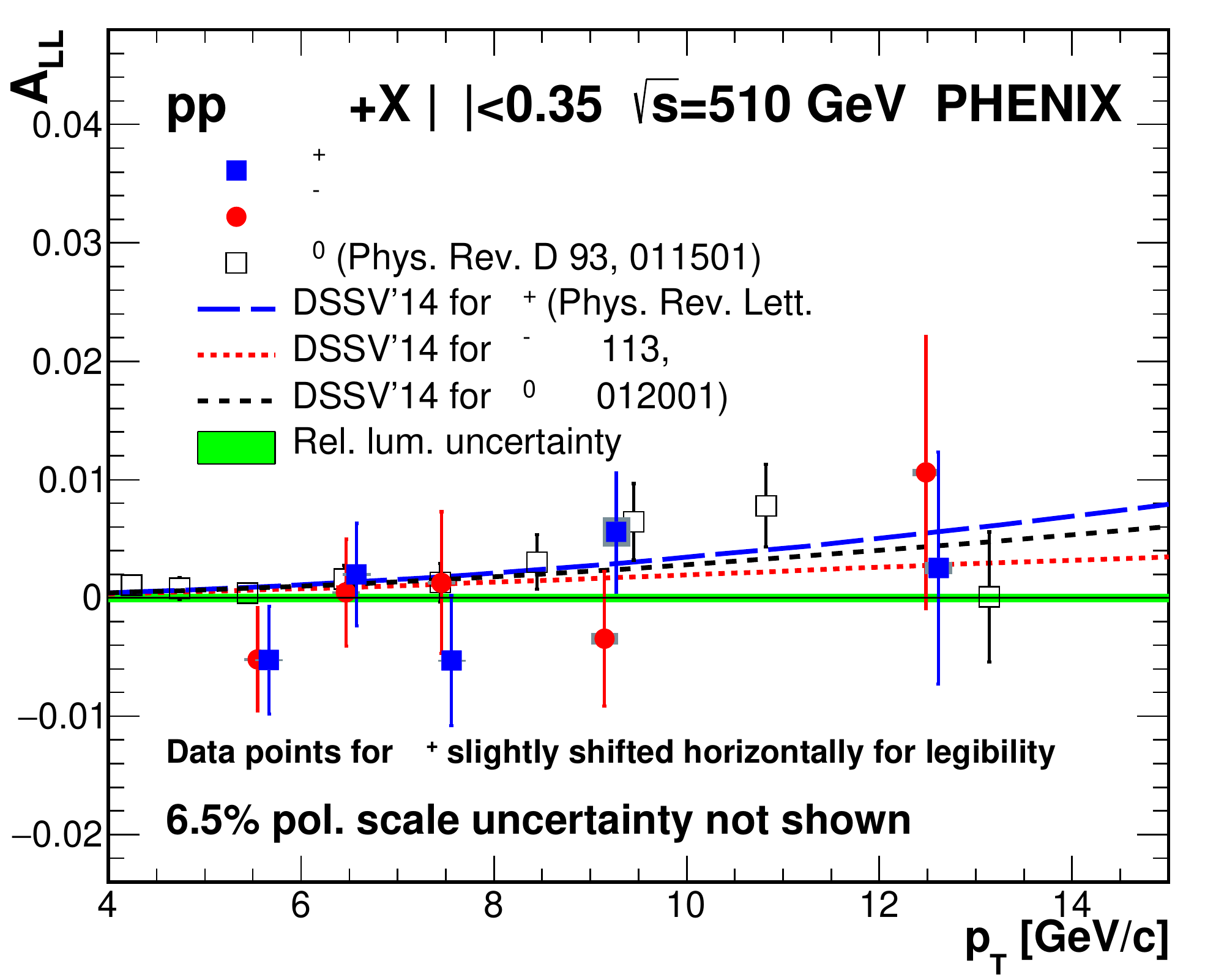}
\caption{\ALL\ vs \pT. \ALL's of previously published neutral pions (open black squares) \cite{PhysRevD.93.011501} and DSSV14 calculations \cite{PhysRevLett.113.012001} are also shown.}
\label{fig:all-pi-pt}
\end{subfigure}
\begin{subfigure}[t]{0.48\textwidth}
\includegraphics[width=\textwidth]{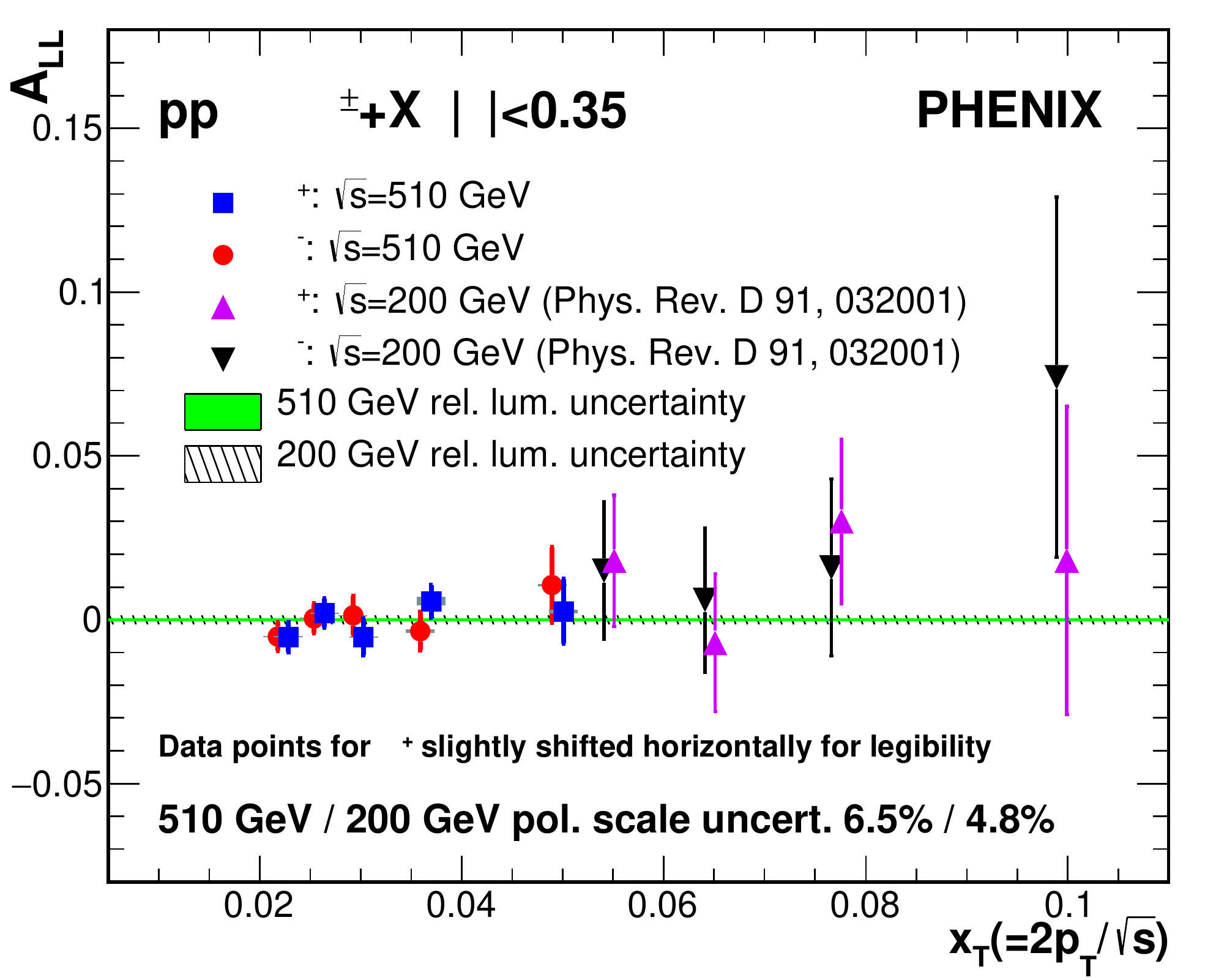}
\caption{\ALL's of charged pions for positive (closed blue squares) and negative pions (closed red circles) at $\sqrt{s}$ = 510 GeV. The statistical uncertainties of asymmetries and the point-to-point systematic uncertainties from background are represented by the continuous lines and the gray bands, respectively.}
\label{fig:all-pi-xt}
\end{subfigure}
\caption{\ALL\ of charged pions.}
\end{figure}

The \ALL's of charged pions are shown in fig.~\ref{fig:all-pi-pt} for \pT\ dependence and fig.~\ref{fig:all-pi-xt} for $x_T$ dependence. The measurements are consistent with calculations from DSSV14 \cite{PhysRevLett.113.012001}. From fig.~\ref{fig:all-pi-xt}, we see that the data from 510 GeV probe lower $x_T$ range. However, there is not enough statistics to decide whether $\pi^+$ or $\pi^-$ has larger \ALL.

\section{Conclusion}

The proton spin decomposition is important to understand the nonperturbative structure of the proton and test various spin sum rules. The gluon spin $\Delta g(x)$ is particularly interesting, but it is not constrained well by existing pDIS and pSIDIS measurements. Longitudinal polarized protons at RHIC are capable of probing the gluon spin at leading order. The direct photon production provides the most ``clean'' probe to the gluon spin due to little fragmentation contributions involved. The jet and charged pions provide complementary probe to the proton spin with more statistics. Together with the PHENIX ongoing forward cluster and forward/central $\eta$ measurements, these measurements will provide new constraints on the gluon spin.

\section*{Acknowledgements}
Thank you to Werner Vogelsang, Sassot Rodolfo and Ignacio Borsa for providing the direct photon cross section and \ALL\ calculations. Thank you to George Sterman for discussion with the principle of maximum conformality (PMC) method.


\paragraph{Funding information}
This work is supported by the Department of Energy and Research Foundation for the State University of New York.




\bibliography{references}

\nolinenumbers

\end{document}